\documentclass[aps,preprint,groupedaddress,onecolumn]{revtex4-1}
\usepackage[english]{babel}
\usepackage{amsmath,amsfonts}
\usepackage[colorinlistoftodos]{todonotes}
\usepackage{braket}

\usepackage{graphics}
\usepackage{graphicx}
\graphicspath{{figures/}}
\usepackage{dcolumn}
\usepackage{bm}
\usepackage{qcircuit}
\usepackage{float}
\usepackage{caption}
\usepackage{subcaption}
\usepackage{color}
\usepackage{pgfplots} 
\pgfplotsset{compat=newest} 
\pgfplotsset{plot coordinates/math parser=false} 

\renewcommand{\eqref}[1]{Eq.~(\ref{#1})}
\newcommand{\figref}[1]{Fig.~\ref{#1}}

\def\firstAuthorLast{Sample {et~al.}} 
\def\Authors{Jun Yang\,$^{1}$, James Brown\,$^{1}$and James Daniel Whitfield\,$^{1}$}
\def\Address{$^{1}$Dartmouth College, Department of Physics and Astronomy, Hanover NH, 03755}
\def\corrAuthor{James Daniel Whitfield}

\def\corrEmail{james.d.whitfield@dartmouth.edu}

\begin{document}

	\title[Measurement on near-term quantum devices]{A comparison of three ways to measure time-dependent densities with quantum simulators} 
	
	\author{\Authors}
	\address{\Address}

	\begin{abstract}
		Quantum algorithms are touted as a way around some classically intractable problems such as the simulation of quantum mechanics.  At the end of all quantum algorithms is a quantum measurement whereby classical data is extracted and utilized.  In fact, many of the modern hybrid-classical approaches are essentially quantum measurements of states with short quantum circuit descriptions.  Here, we compare and examine three methods of extracting the time-dependent one-particle probability density from a quantum simulation: direct $Z$-measurement, Bayesian phase estimation and harmonic inversion.  We have tested these methods in the context of the potential inversion problem of time-dependent density functional theory.  Our test results suggest that direct measurement is the preferable method.
		We also highlight areas where the other two methods may be useful and report on tests using Rigetti's quantum virtual device.  This study provides a starting point for imminent applications of quantum computing.
	\end{abstract}
		\maketitle
	\section{Introduction}
	The real time simulation of quantum systems on a classical computer is a difficult problem even for a supercomputer due to the fact that the Hilbert space grows exponentially with the system size \cite{Georgescu14}. A universal quantum computer is believed to be the solution of the difficulty, where it is known that a wide class of physical systems can be simulated efficiently on a quantum computer 
	\cite{Lloyd1996,Gilyen19,Yudong2018,Georgescu14,berry_efficient_2007,Nielsen02}. But running a practically meaningful quantum algorithm may require a large amount of qubits, e.g. factoring 2048 bit RSA integers may take up to 20 millions qubits\cite{gidney2019factor}, which is far beyond the capacity of the current best 53-qubit quantum computer\cite{arute_quantum_2019}. So the current quantum technology works best when paired with classical algorithms. We have been studying the application of such a hybrid algorithm in quantum chemistry. The primary example is the time-dependent density functional theory  (TDDFT) \cite{PhysRevLett.52.997}. To utilize quantum technology in classical algorithms, quantum measurement are necessary, here we measured the density operator on Rigetti's quantum device and then utilized the density to perform the potential inversion within the framework of TDDFT.
	
	Density functional theory (DFT) is a powerful tool in modeling condensed matter systems \cite{PhysRev.136.B864}. In the  framework of DFT, a non-interacting system with a self-consistently determined potential is constructed to replace the interacting system.  The additional potential term in the non-interacting system is known as the Kohn-Sham potential. Such a system with non-interacting particles is the Kohn-Sham(K-S) system. In the K-S system, the calculation of an exchange-correlation term is required. However, the exact form of the exchange-correlation potential is not yet known. This term is usually obtained with some approximation methods \cite{PhysRev.81.385,PhysRevLett.77.3865,Barth_1972,PhysRevB.7.1912}, machine learning methods \cite{PhysRevLett.108.253002, nagai_completing_2020}. In fact, the utilization of  a quantum computer can help generate an accurate exchange-correlation potential.
	This idea is mentioned in the article \cite{2019arXiv190305550H}, where a hybrid method of generating exchange-correlation potential for classical DFT calculation is proposed.
	
	The time dependent counterpart of DFT, time-dependent density functional theory (TDDFT) is widely used in finding the dynamics of the system when a time-dependent potential is present. Similar to DFT, TDDFT uses the time dependent K-S system where a time-dependent K-S potential is required. We call the task of constructing such a K-S potential when given the time-evolution of the on-site probability density, the K-S potential inversion problem.  In article \cite{Whitfield_2014}, a scheme of solving the K-S potential inversion problem utilizing a quantum computer was proposed. We have recently returned to this proposal with improved numerical methods for inverting the potential \cite{Brown2019}. To obtain the K-S potential, we need to get the density of the time evolved many-particle system using a quantum computer. 
	
	In this paper, we will present three different methods of measuring the density operator on a quantum computer and compare the performance of the methods. 
	
	An outline for the remainder of the article is as follows:
	first, we discuss the phase estimation approach to measurement.  Then we describe the circuit implementation for measuring the on-site fermionic density.  Qubit descriptions for the fermionic operator are explained in the next part followed by  the illustration of a two-electron test.  Finally, three schemes for extracting the density are tested numerically and compared.
	
	\section{Methods}
	
	\subsection{phase estimation}
	
	Quantum phase estimation \cite{Kitaev1995, qpe} plays an important role in the quantum algorithm zoo \cite{zoo}, it is a key sub module of many quantum algorithms \cite{PhysRevLett.103.150502, doi:10.1137/S0097539795293172,365700}. It is also an important procedure to measure the on-site density operator in our work.
	
\begin{figure}[t]
		$$
		\Qcircuit  @C=1em @R=2em {
			\lstick{\ket{0}}&\qw &\qw  &\qw  \barrier[-1.5em]{1}
			&\gate{\sf H}  \qw &\ctrl{1}       & \gate{\sf H} &\qw &\meter \\
			\lstick{\ket{\psi}} &\qw &\gate{U(t)}  &\qw  &\qw  &\gate{U_O(\tau)}     &\qw           &\qw & \rstick{\ket{\psi^\prime}} \qw
		}
		$$
		\caption{The circuit for measuring the density matrix. The half before the dashed line is used for evolving the state to time $t$, the half after is used for doing the  measurement of an observable at time $t$ via phase estimation.}
		\label{dscirc}
	\end{figure}
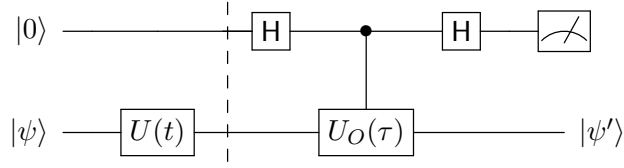

	We'll next describe the general picture of doing the measurement of an arbitrary operator and how quantum phase estimation plays a role in our work. To implement the measurement of an arbitrary observable, we will consider the circuit as shown in \figref{dscirc}.  The circuit has two parts, the part before the dashed line is for evolving the initial state at time $t$ under a fixed fermionic Hamiltonian of chemical interests. 
	
	The system of the most chemical interests is  the interacting electron system. The Hamiltonian of a many-body interacting system is given by
	
	\begin{align}
		H =  \sum_i^N \left[-\frac{\nabla_{i}^2}{2} +V_{ext}(\mathbf{r}_i) +\frac{1}{2}\sum_{j}^{N} \frac{1}{|\mathbf{r}_i-\mathbf{r}_j|}\right]
	\end{align}
	
	where $V_{ext}(\mathbf{r}_i)$ is the external potential energy consists of the interaction between the electrons and the external field. 
	
	The second quantized form of the above many-body Hamiltonian is given by
	
	\begin{align}
		H = \sum_{pq} h_{pq} a^\dagger_p a_q + \frac12 \sum_{pqrs} h_{pqrs} a^\dagger_p a^\dagger_q a_r a_s
	\end{align} 
	where the fermionic operators $\{a_p, a_p^\dagger\}$ satisfy $a^\dagger_q a_p + a_p a^\dagger_q = \delta_{pq}, a_pa_q=-a_qa_p$ and $a_p^\dagger a_q^\dagger=-a_q^\dagger a^\dagger_p$. Given the basis set $\{\chi_p(\mathbf{r})\}$, the coefficients $h_{pq}, h_{pqrs}$ are given by
	
	\begin{align}
		h_{pq} = \int d\mathbf{r}   \chi^\star_p(\mathbf{r})\left(-\frac{1}{2}\nabla^2 + V_{ext}(\mathbf{r})\right) \chi_q(\mathbf{r}) \\
		h_{pqrs} = \int d\mathbf{r}_1 d\mathbf{r}_2 \frac{\chi^\star_p(\mathbf{r}_1) \chi^\star_q(\mathbf{r}_2) \chi_r(\mathbf{r}_2) \chi_s(\mathbf{r}_1)}{|\mathbf{r}_2 - \mathbf{r}_1|}
	\end{align}
	
	The latter half is a phase estimation circuit  where $U_O(\tau) = e^{-i O \tau}$ where $O$ is the observable to be measured.  
	
	For a general state $\ket{\psi}$, we can expand it in the eigenspace of the operator $O$. To be precise, for a general state $\ket{\psi} = \sum_k c_k \ket{k}$, where $O_k$ and $\ket{k}$ are the eigenvalue and eigenvector of the operator $O$. Thus the probability of measuring zero on the top register  is given by
	\begin{align}
		P(0|\tau,t) &= \frac12 + \frac14 [ \bra{\psi(t, \tau)}\psi(t)\rangle+ \bra{\psi(t)}\psi(t,\tau)\rangle]\nonumber\\
		&=\sum_k |c_k(t)|^2\cos^2\left(\frac{O_k \tau}{2}\right)\nonumber\\
		&= \frac{1}{2} + \frac{1}{4}\sum_k |c_k(t)|^2\left(e^{i O_k \tau} + e^{-i O_k \tau}\right)\label{eq:5}
	\end{align}
	
	where $c_k(t) = \braket{k|U(t)|\psi}$.
	
	In this article, we only consider $O =n_j= a_j^\dagger a_j$ in order to measure the local on-site density at site $j$. This is because the inverse potential is determined by the on-site density and its first and second order derivatives \cite{Whitfield_2014, Brown2019}.
	
	The eigenvalues of $n_j=a_j^\dagger a_j$ are 0 and 1, so the wave function after the unitary evolution $U(t)$ is given by $\ket{\psi(t)}=c_0(t) \ket{\psi_{n_j=0}}+c_1(t)\ket{\psi_{n_j=1}}$. Thus the expectation value of the density is given by
	\begin{align}
		\braket{n_j(t)}=\braket{\psi(t)|a_j^\dagger a_j|\psi(t)} = |c_1(t)|^2
	\end{align}
	
	\subsection{Qubit Encoding}

	To implement the evolution and phase estimation algorithm on a quantum computer, we need to encode the Hamiltonian into qubits. A standard way is to use Jordan-Wigner (JW) transformation, which encodes a fermionic system of $M$ orbitals into $M$ qubits.
	
	\begin{align}
		a_p = \frac{1}{2} \left(X_p + i Y_p\right)Z_1Z_2\dots Z_{p-1}\\
		a_p^\dagger = \frac{1}{2} \left(X_p - i Y_p\right)Z_1Z_2\dots Z_{p-1}
	\end{align}

	With the above transformation, the fermionic Hamiltonian can be encoded into qubit representation. Thus the Hamiltonian can be written as $H = \sum_i h_i$, where all the $h_i$'s are tensor product of Pauli operators.
	
	There is not an easy way to construct arbitrary unitary operators on a quantum computer \cite{PhysRevA.61.032312}. A pragmatic way to simulate the propagator $U(t) = e^{-i H t}$ is applying the Trotter decomposition. 
	
	\begin{align}
		U(t)=e^{-i H t} \approx  \left(e^{-i h_1 t/N} e^{-i h_2 t/N}\dots e^{-i h_n t/N}\right)^N
	\end{align}

	Each individual term in the decomposition above can be simulated efficiently on a quantum computer  \cite{whitfield_simulation_2011}. 
	
	\section{Results and discussion}
	
	We tested our methods on two different Hamiltonians, one is the 4-orbital HeH$^+$ model, the other is the 8-orbital HeH$^+$. For the 4-orbital model, the basis set used to examine the HeH$^+$ molecule is that given in Reference~\citenum{szabo2012modern} which results in four spin orbitals. The interatomic distance is 1.401 Bohr. The basis functions are orthogonalized and then transformed such that the one-body Coulomb matrix is diagonal. This transformation was chosen so that a corresponding scalar time-dependent Kohn-Sham potential could be calculated using for this system using the method of Reference~\citenum{Brown2019}.

    For the larger case, we used HeH$^+$ at the same geometry but in the 6-31G basis set \cite{Ditchfield71}.  This resulted in twice the number of basis functions as the minimal example.  The integrals in the 6-31 basis were computed using Pis4 \cite{psi4}.
	
	The basis functions are orthogonalized and then transformed such that the one-body Coulomb matrix is diagonal. This transformation was chosen so that a corresponding time-dependent Kohn-Sham potential could be calculated using for this system using the method of Reference~\citenum{Brown2019}. In both models, the initial state at $t = 0$ places two electrons in the first two modes of opposite spin. This state is obtained by employing two $X$-gates to prepare $\ket{\psi(0)}=\ket{1100}$ in the 4-orbital model and $\ket{\psi(0)}=\ket{11000000}$ in the 8-orbital model. 
	
	Using Rigetti's quantum virtual machine \cite{Forest}, we then evolve the system under its Hamiltonian for times less than three atomic units.  The propagation is implemented via the first-order Trotterization with Trotter step equal to three. To reduce the Trotter error in evolution, either a shorter Trotter step or a higher order Trotter approximation must be used \cite{Hatano2005}.  This means more quantum gates are needed, making it harder to be implemented on a near term device. Additional sources of error are associated with finite sampling from the binomial distribution and the error associated with the inference steps.  To make the virtual machine slightly closer to a real quantum computer, in all methods below, measurement noise was added into the system, giving 1\% probability of flipping the qubit. It should be noted that the quantum noise found on the actual device was much higher, so we will not present the results from the actual quantum device.
	
\begin{figure*}[htbp]
		\hspace*{-2cm}
		\includegraphics[width=1.2\columnwidth]{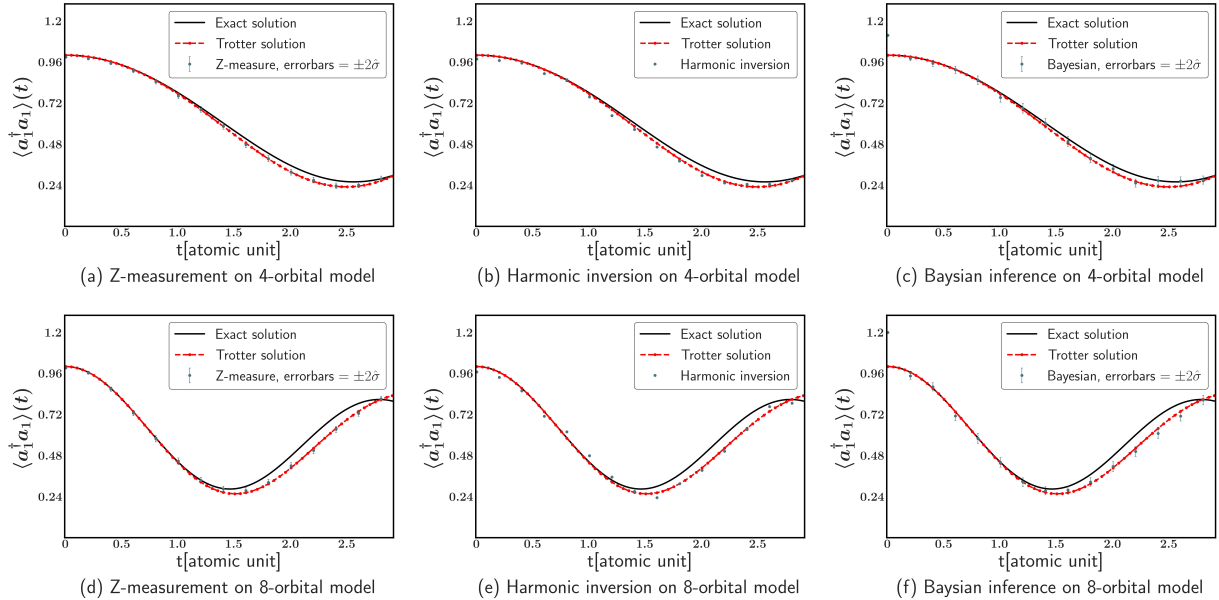}
		\caption{The expectation value of $a_1^\dagger a_1$ is measured via (a) and (d) direct $Z$-basis measurement, (b) and (e) harmonic inversion, and (c) and (f) Bayesian inference. 
			The upper panel are the results of 4-orbital model, the lower panel are the results of the 8-orbital model.
			All of the data points are plotted against the exact Trotter solution depicted in as a continuous line.  In plot (c) and (f), the first point shows a large deviation from the Trotter solution typical of the behavior of Bayesian inference whenever the exact density's value is close to one. Error bars shown in (a), (c) and (d),(f) are  two standard deviations about the mean. The harmonic inversion does not have error bars because the error comes from two sources: from the sampling error at different time $\tau$ and from the reconstruction of the density using harmonic inversion.}
		\label{comp}
	\end{figure*}

	In Figs.~\ref{comp}(a) (d), and \ref{comp}(c) (f), we used 3000 quantum measurement samples per time-point.  For harmonic inversion, a total of 120,000 quantum measurements occur for extracting the density at each time point.   This is because there were $40$ equally spaced $\tau$-points and 3000 quantum measurements were used per fixed $\tau$. A comparison between the measuring results, the exact solution of the original Hamiltonian and the exact solution of the Trotterized Hamiltonian are compared in Fig.~\ref{comp}. In each subfigure, the dark green dots are the result from measurement result, the red dashed line is the solution of the Trotterized Hamiltonian, the black solid line is the solution of the original Hamiltonian.
	
	\subsection{Method 1: Z-basis Measurement}
	
	In the first method, we rely on the fact that
	the Jordan-Wigner transformation of the on-site density operator has a simple form  $a_p^\dagger a_p = (1 - Z_p)/2$.  Thus, we can directly measure the local density operator by measuring $Z_p$ without passing in the phase estimation circuit after the dotted line in \figref{dscirc}.
	
	For an arbitrary wave  function  $\ket{\psi(t)} = c_0(t) \ket{\psi_{{n_p=0}}} + c_1(t) \ket{\psi_{{n_p=1}}}$, where  $\psi_{n_p}$ denotes the state projected into the subspace where the $p$-th qubit is in state $n_p$. Given the fact that $\braket{Z_p(t)} =|c_0(t)|^2 -|c_1(t)|^2$ and $|c_0(t)|^2 +|c_1(t)|^2 = 1$, both amplitudes $|c_0(t)|^2$ and $|c_1(t)|^2$ can be obtained from the measurement.
	
	By repeating the measurement at each time step in the time range $0 \leq t \leq 3$, we obtain the expectation value of the density.   The results based on 15 equally spaced time-points with 3000 measurements at each fixed time are shown in \figref{comp}(a)(d).  The exact time evolution of the density is also shown in the figure for comparison along with error bars of $2\sigma$ reflective of the $N=3000$ sample variance of the binomial distribution.
	
	The simplicity of this measurement approach reduces the classical runtime to the lowest of the three methods compared, and the convergence of the error bars is faster than the Bayesian measurement discussed later.
	
	\subsection{Method 2: Harmonic Inversion}
	
	Harmonic inversion is a technique of extracting the amplitudes $A_j$, frequencies $f_j$, phases $\phi_j$ and exponential decay constants $\alpha_j$ out of a signal,
	
	\begin{align}
		f(\tau) = \sum_j A_j e^{-i (2\pi f_j \tau - \phi_j)  - \alpha_j \tau}
	\end{align}

	which is evenly sampled \cite{harminv,Mandelshtam2001}. The signal reconstructed from harmonic inversion has the same form as the probability $P(0|\tau)$ except for the decaying term which is negligible when the decoherence is not considered.   By comparing the form of the reconstructed signal with the probability, we can obtain the density from the reconstructed signal.
	
	The results of density measurement through harmonic inversion are shown in \figref{comp}(b)(e). Each point in \figref{comp}(b)(e) was computed through harmonic inversion using the HarmInv package \cite{harmInvPackage}. Because the local density operator $a_p^\dagger a_p$ only has eigenvalues zero and one, the measurement outcome has a simple form
	
	\begin{align}
		P(0|\tau, t) = A_0(t) + A_1(t) \left(e^{-i 2\pi f \tau} + e^{i 2\pi f \tau}\right)
	\end{align}

	where $A_0(t)=\frac12( 2-|c_1(t)|^2)$, $A_1(t)=|c_1(t)|^2/4$, and $f=1/2\pi$. 
	
	\subsection{Method 3: Bayesian inference}
	
	Bayesian inference can be used to estimate the density as well. As a powerful tool of making inferences, Bayesian inference has wide applications. We applied Bayesian inference to infer the unknown parameters in a quantum system which, in our case, is the on-site density. The density estimation was implemented via sequential Monte Carlo (SMC) \cite{Granade_2012}. This method requires the most communication between the classical and quantum processors since the SMC suggests each $\tau$-point based on the previous outcomes.
	The Bayesian experimental design is based on the implementation found in the QInfer package \cite{Granade2017qinferstatistical}. Bayesian inference gives the probability distribution of a parameter over the parameter space.  The final decision is made according to the posterior probability $P(\theta| d_1, d_2 \dots d_N)$, where $\theta$ is the parameter we want to estimate, $d_i$'s are the outcome of each measurement.  In the present application, $\theta\equiv \langle n_j(t)\rangle$.
	
	Recall the Bayesian rule, the posterior probability is updated by carrying out experiments sequentially,
	
	\begin{align}
		P(\theta | d_1, d_2, \dots d_N) \propto \prod_{i=1}^{N}P(d_i|\theta)P(\theta)
	\end{align}

 where $P(\theta)$ is the prior probability, $P(d_i|\theta)$ is the likelihood function.
	
	The likelihood function contains the information about the parameters before conducting any experiments. Since we know nothing before the experiment, we can initialize the prior with a uniform distribution over the parameter space. For the phase estimation circuit of \figref{dscirc}, the likelihood function is given by 
	
	\begin{align}
		P(d | \braket{n_j(t)}; \tau) = \frac{1}{2} + \frac{(-1)^d}{4} \braket{\psi(t)|  \{U_O(\tau) + U_O^\dagger(\tau)\}|\psi(t)}
	\end{align}

	where  $U_O(\tau)=\exp(-i\tau a_j^\dagger a_j)$ and $d = 0 \text{ or } 1$.  Note, when $d=0$ we recover \eqref{eq:5}.

	With this we can rewrite the likelihood function as 
	
	\begin{align}
		P(d | \braket{n_j(t)}; \tau)  =  \delta_{d,0} + \frac{(-1)^d}{2}(\cos \tau - 1) \braket{n_j(t)}
	\end{align}

	This can be compared with \eqref{eq:5} in the case that $d=0$.
	
	The results of Bayesian inference are shown in \figref{comp}(c)(f). Bayesian inference has good performance within a wide range of the time domain except at the boundary of the estimate domain e.g. when the density is one or zero.  This is based on numerical evidence since the majority of the points at or near the boundary of the estimation domain needed to be discarded when cleaning the data as discussed below.
	
	Unlike harmonic inversion, $\tau$ in the phase estimation circuit is not required to be evenly spaced. Another advantage of Bayesian inference is that we do not need to know the exact form of the function to be estimated \emph{a priori}. Bayesian inference could also be applied to estimate more general parameters.
	
	\emph{Comparison}
	
	To quantify the accuracy of these density extraction methods, we employ the $L_1$ norm to measure the deviation from the Trotter solution. For discrete data points, the deviation is given by the loss function on the density at the first site: $L = \sum^N_{k} |\Tilde{n}_1(t_k) - n_1(t_k))| / N$, where $\Tilde{n}_1(t_k)$ is the outcome of the measurement at time $t_k$, $n_1(t_k)$ is the solution of the Trotterized Hamiltonian.
	
	\begin{figure*}[tbh]
		\includegraphics[width=\columnwidth]{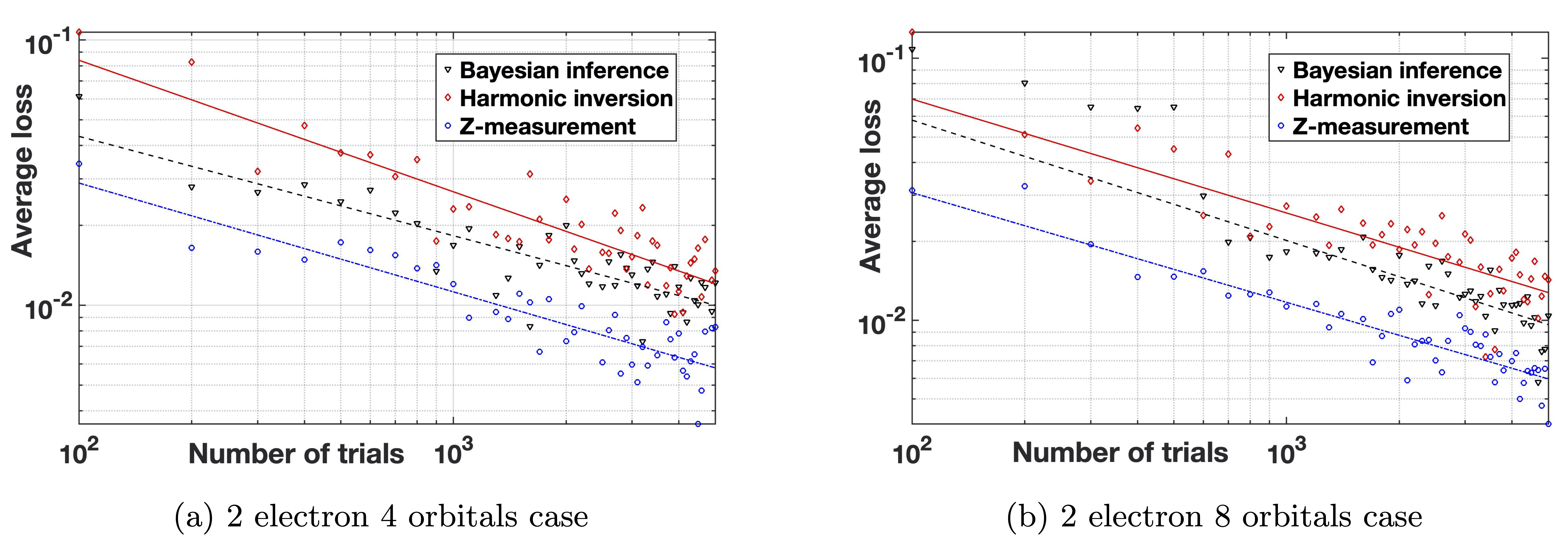}
		\caption{Loss function versus number of measurements of the quantum computer (trials) in both 4-orbital model and 8-orbital model. Red diamonds, black triangles and blue circles are drawn from harmonic inversion, Bayesian inference and $Z$-basis measurement respectively. }
		\label{fig:pf}
	\end{figure*} 
	
	\figref{fig:pf} shows how the loss function scales with the number of trials for each of the three approaches.  The convergence rate for determining the bias of a coin would be 0.5 but here additional measurement error has been introduced into the model which prevents $L=0$ situation even with an infinite number of samples.  Further, in our implementation, the Bayesian and harmonic inversion techniques sometimes reported anomalously poor estimates of the density at a given time.  A single fluctuation of this type along the time trace of the density entirely dominates the loss function. For the sake of comparison, we did not include the data points that are $5\sigma$ away from the exact solution in all three methods. This led to more stable results when the number of trials is small. Another benefit of filtering the data is that for the Bayesian inference, estimates close to the boundary of the domain are subject to large fluctuations giving poor estimates. So we can exclude the wrong data points by setting a $5\sigma$ window. Although the discarding procedure is \emph{ad hoc} and requires knowing the exact answer, we have tested our data at various levels of cutoff finding that at any fixed cutoff harmonic inversion had the most points discarded and consistently displayed marginally faster convergence rates. 
	
	Fig.~\ref{fig:pf}(a) shows the scaling of loss function  of the 4-orbital model. The slope of the fitting lines are  -0.4961, -0.3747 and -0.4103 respectively. Points  $5\sigma$ away from the exact density under the Trotter approximation are not used for calculating the loss function.  This resulted in  73.87\%, 87.26\% and 90.07\% of points used in the plotted data respectively. 
	
	Fig.~\ref{fig:pf}(b) shows the scaling of loss function of the 8-orbital model. The slope of the fitting lines are  -0.4346, -0.4595 and -0.4184 respectively. Points  $5\sigma$ away from the exact density under the Trotter approximation are not used for calculating the loss function.  This resulted in 80.53\%, 84.40\% and 93.20\% of points used in the plotted data respectively.
	
	Harmonic inversion measures 40 times more than the other two methods, so the actual data and fitting line should be shifted to the right by 40 times the number of measurements showing in the figure.
	
	Regardless of the possible improvement in convergence, it should be reminded that the harmonic inversion technique uses many quantum computer queries to estimate $P(0|\tau,t)$ at variable $\tau$ before inferring the density at a fixed time $t$. In comparing the three methods, all require time evolution of the system wave function to time $t$.  In the harmonic inversion and Bayesian estimation techniques, additional gates are needed for the $\tau$ propagation under the observable for density.  The difference between queries in harmonic inversion and Bayesian inference is the selection of the $\tau$ parameter in $U_O(\tau)$.
	
	While the convergence rates are all approximately the same, it is clear that the $Z$-basis measurement has the best performance in terms of the number of queries of the quantum computer.  In the case considered here, the direct $Z$ measurements are convenient for the Jordan-Wigner encoding.  In other circumstances with different fermion-to-spin transforms, the direct measurement technique may not be as fruitful. 
	For existing and near-term quantum devices, the constraints of low circuit depth suggests direct measurement of the $Z$ operators as the best path forward when using a Jordan-Winger transformed qubit Hamiltonian. 
	
	The runtime of these three methods also varies.  Since direct $Z$-measurements are the simplest from an inference point of view, the classical computation time is also the least.  Bayesian inference requires many steps for the sequential Monte Carlo to converge \cite{Granade_2012}.  Consequently, this method used the longest amount of classical computational time.  Although harmonic inversion uses 40 times more measurement per time-point, it is interesting to note that it only took an intermediate amount of classical processing time.
	
	\section{Conclusions}
	
	We have tested three different methods of measuring the on-site density operator for a toy model inspired by TDDFT.  We were able to conclude that direct $Z$ measurements obtains the best estimates of the on-site density for a given number of quantum computer queries.  This is based on the use of the Jordan-Wigner transform and simulated measurement noise.  Of course, we could have considered other fermion-to-spin transforms which lead to different encodings of the $a_i^\dagger a_i$.  
	
	For improving our noise models, we can do no better than testing our circuits on current and future quantum devices.  We tested our circuits on Rigetti's quantum device but found that the loss function depends heavily on which qubits are used as well as the permutation of qubit labels within the circuit.  Time evolution under the full Hamiltonian did not return any signal even when using only one first-order Trotter step. We therefore resorted to using a truncated Hamiltonian which included the one-body Hamiltonian and only the Coulomb-like ($h_{ijji}$) terms of the two-body Hamiltonian. After encoding and exponentiation, this Hamiltonian results in 66 universal gates and compiled non-deterministically using the PyQuil package \cite{Forest} to approximately 200 allowable gates on the Rigetti device. Due to decoherence, only a weak signal was present where amplitudes recovered were between three and twenty percent of the exact solution.  The recovered amplitude depended mostly on qubit selection but also changed run-to-run. The frequency and sinusoidal shape of the signal was recovered more reliably. In our present study, the eigenenergies were not interesting but we suspect that problems that depend on the frequencies may be more successfully calculated on the current Rigetti device.
	
	We plan to continue our inquiry into the TDDFT potential inversion problem using existing and forthcoming quantum technology.  Tasks that avoid QMA-hard\cite{QMA} state preparation problems will continue to be of interest to those looking for new applied areas of quantum computation.
	
	\section{Acknowledgements}
	JY, JB and JDW were supported by the U.S. Department of Energy, Office
	of Science, Office of Advanced Scientific Computing Research, under the Quantum
	Computing Application Teams program (Award 1979657).  JDW was also supported 
	by the NSF (Grant 1820747) and 
	additional funding from the DOE (Award A053685). Calculations were performed using Dartmouth's Discovery Linux HPC Cluster.
	\bibliographystyle{frontiersinSCNS_ENG_HUMS}
	\bibliography{pea.bib}{}
\end{document}